
%
%

\newcount\xheure
\newcount\xminute
\newcount\xaux
\def\heure{
\xminute=\time
\xheure=\xminute
\divide\xheure by 60
\xaux=\xheure
\multiply\xaux by -60
\advance\xminute by \xaux
\ifnum \xminute<10 \def\zero{0} \else \def\zero{} \fi
\number\xheure h\zero\number\xminute\ }
\hsize=6.0in
\hoffset=0.1875in
\tolerance=1000\hfuzz=2pt
\baselineskip=14.4pt
\newcount\yearltd\yearltd=\year\advance\yearltd by -1900
\def\title#1#2{
   \vsize=8.0in\voffset=0.5in
   \nopagenumbers\pageno=0\abstractfont
   \rightline{}
   \rightline{}
   \vskip 0.8in plus 0.3in
   \centerline{\titlefont\bf #1}
   \vskip 0.2in
   \centerline{\titlefont\bf #2}
   \vskip 0.38in plus 0.1in}

\def\finishtitlepage#1{\vfill
   \leftline{#1}\tenpoint\supereject\global
   \footline={\hss\tenrm\folio\hss}
   \baselineskip=18pt
   \vsize=8.9in\voffset=0.5in}

\def\date#1{\finishtitlepage{#1}}

\def\nolabels{\def\eqnlabel##1{}\def\eqlabel##1{}\def\figlabel##1{}%
        \def\reflabel##1{}}
\def\writelabels{\def\eqnlabel##1{%
        {\escapechar=` \hfill\rlap{\hskip.11in\string##1}}}%
        \def\eqlabel##1{{\escapechar=` \rlap{\hskip.11in\string##1}}}%
        \def\figlabel##1{\noexpand\llap{\string\string\string##1\hskip.66in}}%
        \def\reflabel##1{\noexpand\llap{\string\string\string##1\hskip.37in}}}
\nolabels
\global\newcount\secno \global\secno=0
\global\newcount\meqno \global\meqno=1
\global\newcount\ssecno \global\ssecno=0

\def\newsec#1{\global\advance\secno by1 \global\ssecno=0
\xdef\secsym{\the\secno.}\global\meqno=1
\bigbreak\bigskip
\noindent{\bf\the\secno. #1}\par\nobreak\medskip\nobreak}
\xdef\secsym{}

\def\newssec#1{\global\advance\ssecno by1
\bigbreak\bigskip
\indent{\bf\the\secno.\the\ssecno. #1}\par\nobreak\medskip\nobreak}

\def\appendix#1#2{\global\meqno=1\xdef\secsym{\hbox{#1.}}\bigbreak\bigskip
\noindent{\bf Appendix #1. #2}\par\nobreak\medskip\nobreak}


\def\eqnn#1{\xdef #1{(\secsym\the\meqno)}%
\global\advance\meqno by1\eqnlabel#1}
\def\eqna#1{\xdef #1##1{\hbox{$(\secsym\the\meqno##1)$}}%
\global\advance\meqno by1\eqnlabel{#1$\{\}$}}
\def\eqn#1#2{\xdef #1{(\secsym\the\meqno)}\global\advance\meqno by1%
$$#2\eqno#1\eqlabel#1$$}

\def\seqnn#1#2{\global\advance\meqno by-1\xdef
#1{(\secsym\the\meqno')}\global\advance\meqno by1%
$$#2\eqno#1\eqlabel#1$$}

\def\myfoot#1#2{{\baselineskip=14.4pt plus 0.3pt\footnote{#1}{#2}}}
\global\newcount\ftno \global\ftno=1
\def\foot#1{{\baselineskip=14.4pt plus 0.3pt\footnote{$^{\the\ftno}$}{#1}}%
\global\advance\ftno by1}

\global\newcount\refno \global\refno=1
\newwrite\rfile

\def\ref{[\the\refno]\nref}
\def\nref#1{\xdef#1{[\the\refno]}\ifnum\refno=1\immediate
\openout\rfile=refs_tmp\fi\global\advance\refno by1\chardef\wfile=\rfile
\immediate\write\rfile{\noexpand\item{#1\ }\reflabel{#1}\pctsign}\findarg}
\def\findarg#1#{\begingroup\obeylines\newlinechar=`\^^M\passarg}
{\obeylines\gdef\passarg#1{\writeline\relax #1^^M\hbox{}^^M}%
\gdef\writeline#1^^M{\expandafter\toks0\expandafter{\striprelax #1}%
\edef\next{\the\toks0}\ifx\next\null\let\next=\endgroup\else\ifx\next\empty%
\else\immediate\write\wfile{\the\toks0}\fi\let\next=\writeline\fi\next\relax}}
{\catcode`\%=12\xdef\pctsign{

\def\addref#1{\immediate\write\rfile{\noexpand\item{}#1}} 

\def\listrefs{\vfill\eject\immediate\closeout\rfile
     \centerline{{\bf References}}\bigskip{\frenchspacing%
     \catcode`\@=11\escapechar=` %
     \input refs_tmp\vfill\eject}\nonfrenchspacing}

\def\startrefs#1{\immediate\openout\rfile=refs_tmp\refno=#1}

%
\global\newcount\figno \global\figno=1
\newwrite\ffile
\def\fig{\the\figno\nfig}
\def\nfig#1{\xdef#1{\the\figno}\ifnum\figno=1\immediate
\openout\ffile=figs_tmp\fi\global\advance\figno by1\chardef\wfile=\ffile
\immediate\write\ffile{\medskip\noexpand\item{Fig.\ #1:\ }%
\figlabel{#1}\pctsign}\findarg}
\def\listfigs{\vfill\eject\immediate\closeout\ffile{\parindent48pt
\baselineskip16.8pt\centerline{{\bf Figure Captions}}\medskip
\escapechar=` \input figs_tmp\vfill\eject}}


\font\titlerm=cmr10 scaled \magstep3
\font\titlerms=cmr10 scaled \magstep1

\font\titlebf=cmbx10 scaled \magstep3
\font\titlebfs=cmbx10 scaled \magstep1

\font\titlei=cmmi10 scaled \magstep3  
\font\titleis=cmmi10    scaled \magstep1

\font\titlesy=cmsy10 scaled \magstep3   
\font\titlesys=cmsy10 scaled \magstep1

\font\titleit=cmti10 scaled \magstep3   

\skewchar\titlei='177 \skewchar\titleis='177 
\skewchar\titlesy='60 \skewchar\titlesys='60 

\def\titlefont{\def\rm{\fam0\titlerm}
   \textfont\bffam=\titlebf \def\bf{\fam\bffam\titlebf}
   \textfont0=\titlerm \scriptfont0=\titlerms 
   \textfont1=\titlei  \scriptfont1=\titleis  
   \textfont2=\titlesy \scriptfont2=\titlesys 
   \textfont\itfam=\titleit \def\it{\fam\itfam\titleit} \rm}


\font\tenrm=cmr10 scaled \magstep1
\font\sevenrm=cmr7 scaled \magstep1
\font\fiverm=cmr5 scaled \magstep1

\font\tenbf=cmbx10 scaled \magstep1
\font\sevenbf=cmbx7 scaled \magstep1
\font\fivebf=cmbx5 scaled \magstep1

\font\teni=cmmi10 scaled \magstep1
\font\seveni=cmmi7 scaled \magstep1
\font\fivei=cmmi5 scaled \magstep1

\font\tensy=cmsy10 scaled \magstep1
\font\sevensy=cmsy7 scaled \magstep1
\font\fivesy=cmsy5 scaled \magstep1

\font\tenex=cmex10 scaled \magstep1
\font\tentt=cmtt10 scaled \magstep1
\font\tenit=cmti10 scaled \magstep1
\font\tensl=cmsl10 scaled \magstep1

\def\tenpoint{\def\rm{\fam0\tenrm}
        \textfont0=\tenrm \scriptfont0=\sevenrm \scriptscriptfont0=\fiverm
        \textfont1=\teni  \scriptfont1=\seveni  \scriptscriptfont1=\fivei
        \textfont2=\tensy \scriptfont2=\sevensy \scriptscriptfont2=\fivesy
        \textfont\itfam=\tenit \def\it{\fam\itfam\tenit}
        \textfont\ttfam=\tentt \def\tt{\fam\ttfam\tentt}
        \textfont\bffam=\tenbf \def\bf{\fam\bffam\tenbf}
        \textfont\slfam=\tensl \def\sl{\fam\slfam\tensl} \rm
   \setbox\strutbox=\hbox{\vrule height 10.2pt depth 4.2pt width 0pt}
   \parindent=24pt\parskip=0pt plus 1.2pt
   \topskip=12pt\maxdepth=4.8pt
   \jot=3.6pt\normalbaselineskip=14.4pt\normallineskip=1.2pt
   \abovedisplayskip=13pt plus 3.6pt minus 5.8pt
   \belowdisplayskip=13pt plus 3.6pt minus 5.8pt
   \abovedisplayshortskip=-1.4pt plus 3.6pt
   \belowdisplayshortskip=13pt plus 3.6pt minus 3.6pt
   \topskip=12pt \splittopskip=12pt
   \scriptspace=0.6pt\nulldelimiterspace=1.44pt\delimitershortfall=6pt
   \thinmuskip=3.6mu\medmuskip=3.6mu plus 1.2mu minus 1.2mu
   \thickmuskip=4mu plus 2mu minus 1mu
   \smallskipamount=3.6pt plus 1.2pt minus 1.2pt
   \medskipamount=7.2pt plus 2.4pt minus 2.4pt
   \bigskipamount=14.4pt plus 4.8pt minus 4.8pt
   \hfuzz=1pt\vfuzz=1pt
   }

\def\abstractfont{\tenpoint}

\tenpoint


\def\noblackbox{\overfullrule=0pt}
\def\dup{~{\vphantom{1}}}
\def\boxeqn#1{\vcenter{\vbox{\hrule\hbox{\vrule\kern3.6pt\vbox{\kern3.6pt
        \hbox{${\displaystyle #1}$}\kern3.6pt}\kern3.6pt\vrule}\hrule}}}
\def\mbox#1#2{\vcenter{\hrule \hbox{\vrule height#2.4in
        \kern#1.2in \vrule} \hrule}}  
\def\CA{\hbox{{$\cal A$}}} \def\CC{\hbox{{$\cal C$}}}
\def\CQ{\hbox{{$\cal Q$}}} \def\CE{\hbox{{$\cal E$}}}
\def\CM{\hbox{{$\cal M$}}} \def\CN{\hbox{{$\cal N$}}}
\def\CV{\hbox{{$\cal V$}}} \def\CW{\hbox{{$\cal W$}}}
\def\CX{\hbox{{$\cal X$}}} \def\CY{\hbox{{$\cal Y$}}}
\def\CF{\hbox{{$\cal F$}}} \def\CG{\hbox{{$\cal G$}}}
\def\CL{\hbox{{$\cal L$}}} \def\CH{\hbox{{$\cal H$}}}
\def\CI{\hbox{{$\cal I$}}} \def\CU{\hbox{{$\cal U$}}}
\def\CB{\hbox{{$\cal B$}}} \def\CZ{\hbox{{$\cal Z$}}}
\def\CR{\hbox{{$\cal R$}}} \def\CO{\hbox{{$\cal O$}}}
\def\CD{\hbox{{$\cal D$}}} \def\CT{\hbox{{$\cal T$}}}
\def\CP{\hbox{{$\cal P$}}} \def\CJ{\hbox{{$\cal J$}}}
\def\CS{\hbox{{$\cal S$}}}
\def\lform{\hbox{$\sqcup$}\llap{\hbox{$\sqcap$}}}
\def\darr#1{\raise1.8ex\hbox{$\leftrightarrow$}\mkern-19.8mu #1}
\def\half{{\textstyle{1\over2}}} 
\def\roughly#1{\ \lower1.5ex\hbox{$\simOw$}\mkern-22.8mu #1\,}
\def\ie{i.e.}
\def\simti{_{ \ \sim \ \atop t \rightarrow \infty }}
\def\equiinf{\lower 0.12cm\hbox{$\widetilde{_{_{t\rightarrow \infty}}}$}}
\def\etauinf{\lower 0.12cm\hbox{$\widetilde{_{_{\tau\rightarrow \infty}}}$}}

\hyphenation{di-men-sion di-men-sion-al di-men-sion-al-i-ty
     di-men-sion-al-ly}
\nolabels
\def\pt{P(t)}
\def\equiinf{\lower 0.12cm\hbox{$\widetilde{_{_{t\rightarrow \infty}}}$}}
\def\equiinfa{\lower 0.12cm\hbox{$\widetilde{_{_{a\tau=t\rightarrow
\infty}}}$}}
\def\mbt{ e^{-m_bt}}
\def\fb{$f_{_B}$}
\def\xequn{$x = 1$}
\def\mbzt{ e^{-m_b^0t}}
\def\atozero{$a \rightarrow 0$}
\rightline{CERN-TH. 6599/92}
\rightline{LPTHE-\number\yearltd/45}
\rightline{ROM2F/92/43}
\vskip -0.5 truecm
 \title{A rigorous treatment of}{ the lattice renormalization problem of $f_B$}
\centerline{ Ph. Boucaud{$^1$}, J.P. Leroy{$^1$}, J. Micheli{$^1$},
O. P\`ene{$^{1,2}$} and
   G.C. Rossi{$^3$}}
\vskip 0.3in
\centerline{{$^1$})
Laboratoire de Physique Th\'eorique et Hautes Energies
 \myfoot{$^{\dagger}$}{Laboratoire associ\'e au CNRS}}
\centerline{Universit\'e de Paris-Sud, F-91405 Orsay, France\myfoot{}{}}
\centerline{{$^2$}) CERN Theory Division, CH-1211 Geneva 23, Switzerland}
\centerline{{$^3$}) Dipartimento di Fisica, Universit\`a di Roma II, Tor
Vergata}  \centerline{Via della Ricerca Scientifica, I-00133 Roma, Italy
\myfoot{ }{CERN-TH.6599/92}}
\centerline{and INFN, Sezione di Roma, Tor Vergata
\myfoot{ }{July 1992}}

\vskip 0.3in
\centerline{\bf Abstract}
\noindent

The $B$-meson decay constant can be measured on the lattice using a $1/m_b$
expansion. To relate the physical quantity to Monte Carlo data one has to
know the renormalization coefficient, $Z$, between the lattice operators
and their continuum counterparts. We come back to this computation to resolve
discrepancies found in previous calculations. We define and discuss in
detail the renormalization procedure that allows the (perturbative)
computation of $Z$. Comparing the one-loop calculations in the effective
Lagrangian approach with the direct two-loop calculation of the two-point
$B$-meson correlator in the limit of large $b$-quark mass, we prove that
the two schemes give consistent results to order $\alpha_s$.
We show that there is, however, a
renormalization prescription ambiguity that can have sizeable numerical
consequences. This ambiguity can be resolved in the framework of an $O(a)$
improved calculation, and we describe the correct prescription in that case.
Finally we give the numerical values of $Z$ that correspond to the different
types of lattice approximations discussed in the paper.
\finishtitlepage{}

\voffset 0.5 cm
\newsec{Introduction}
The study of heavy flavours has attracted a lot
of interest in the last few years, both from the experimental and
theoretical point of view. Many interesting properties of matrix
elements can be derived on rather general
assumptions, using the spin symmetry that emerges in the limit in which the
mass of the heavy quark goes to infinity \ref\EICHOLD{E.
Eichten, Nucl. Phys. {\bf B} (Proc. Suppl.) {\bf 4} (1988) 140.},
\ref\HEFT{ M.B. Voloshin and M.A. Shifman,
Sov. J. Nucl. Phys. {\bf 45} (1987) 292 and {\bf 47} (1988)
511;\hfill\break H.D. Politzer and M. Wise, Phys. Lett. {\bf B 206} (1988) 681
and {\bf B 208} (1988) 504;\hfill\break P. Lepage and B. Thacker, Nucl. Phys.
{\bf B} (Proc.Suppl.) {\bf 4}
(1988) 199;\hfill\break H. Georgi, Nucl. Phys {\bf B361} (1991),
339;\hfill\break J.D. Bjorken, Proc. 25th Int Conf on HEP, Singapore, 1990
(World Scientific, Singapore, (1991), p.1195; Proc. 18th SLAC Summer Inst on
Particle Physics, Stanford, 1990 (SLAC-Report-378, Stanford, 1991),
p.167;\hfill\break N. Isgur and M. Wise, Phys. Lett. {\bf B 232} (1990) 113;
Phys. Lett. {\bf B 237} (1990) 527; Phys. Rev. Lett. {\bf 66} (1991) 1130;
\hfill\break H. Georgi and F. Uchiyama, Phys. Lett. {\bf B 238} (1990) 395;
\hfill\break H. Georgi and M. Wise, Phys. Lett. {\bf B 243} (1990) 279.}.

 Beyond symmetry arguments, the only method based on first principles
that can be used
to predict physical quantities, such
as the $B$-meson decay constant, or the universal form factor
for current matrix elements between heavy mesons,
is lattice QCD, complemented with a
 \def\surm{$1/m_b\ $} \surm expansion of the heavy-quark
propagator, as proposed in ref. \EICHOLD.

In this scheme, the heavy-quark field is no longer explicitly
present on the lattice, but it is replaced by a static colour source,
 and one has only to
consider correlation functions with gauge and light-quark fields.
The latter ones can be evaluated in lattice QCD
by Monte Carlo simulations in the standard way.
This method is the only one applicable to the $B$ meson since
 the $b$ quark is too heavy to live even on lattices with
the smallest lattice spacing presently attainable in numerical
simulations ($a^{-1}$ of the order of 3 to 4 GeV).
In the phenomenological analysis, one then has to
interpolate between the infinite mass limit computed in
this way and the results obtained with the standard approach
for quarks lighter than $a^{-1}$ \ref\ASMAA{A. Abada, C.R. Allton, Ph. Boucaud,
D.B. Carpenter, M. Crisafulli, J. Galand, S. Gusken, G. Martinelli, O.
P\`ene, C.T. Sachrajda, R. Sarno, K. Schilling and R. Sommer, Nucl. Phys. {\bf
B
376} (1992) 172.}.

  From now on we will concentrate on the method based on the infinite mass
limit. We will call it for short in the following the {\it static}
quark approach.

As usual, we have to deal with the problem of lattice renormalization
to relate numbers effectively measured on the lattice to the
corresponding physical quantities. Two methods have been used
 for the calculation of the necessary
 renormalization constants \ref\LIN{Ph. Boucaud, L.C. Lin and O.
P\`ene, Phys. Rev. {\bf D 40} (1989) 1529 and {\bf D 41}
(1990) 3541(E).}, \ref\EICH{E. Eichten
and B. Hill, Phys. Lett. {\bf 240 B} (1989) 193.}.
Apparently the two approaches have led to different numerical results.
One of the purposes of this paper is to understand and to solve this
discrepancy.

Let us focus on the computation of\def\FB{$f_{_B}$} \FB,
the decay constant of a $B$-meson.  From the current-current correlation
function:
\eqn\PT{
P(t) =\int d {\bf x}  \ \langle  J_{qb}^\dagger({\bf 0},0)
J_{qb}{\bf x},t)\rangle  ,
}
where
\eqn\Jqb{
J_{qb}(x)\ =\ \bar b(x) \gamma_{0}\gamma _{5}q(x) ,
}
we can extract the value of \FB\ by looking at its large $t$ behaviour:
\eqn\PTLIM{
 \eqalign  { { P(t)} \ & \equiinf \ {1 \over
{2M_B}} \vert <0 \vert J_{qb}({\bf 0},0)\vert \bar B > \vert ^2
e^{-M_Bt} \,,\cr
& = \ \ {f_{_B}^2M_B\over 2} { e^{-M_Bt}}\,\,.\cr}
}
where $M_B$\ is the $B$-meson mass.

The presence of the heavy $b$ field prevents a direct lattice evaluation of the
expectation value in \PT. The nice idea of \hbox{ref. $\EICHOLD$}\ is to
replace, in the limit
$m_b \to \infty$, the $b$ and  $\bar b$ fields in eq. \PT\ by the static
$b$ propagator:
\eqn\SB{
S_b({\bf x'}, 0; {{\bf x},t})=
\CP_{\bf x}{t\choose 0}\delta({\bf x}-{\bf x'})\left \lbrack \theta(t)\
\mbt{{1- \gamma_{0}} \over 2}  +
\theta(-t)\ { e^{m_bt} }{{1+ \gamma_{0}} \over 2}\right \rbrack
\,\,\,,
}
where $ m_b $\ is the  $b$-quark mass\foot{As we shall see, we do not need
a precise definition of $m_b$.}\  and
\eqn\PY{
\CP_{\bf x}{t{_2}\choose t{_1}}=P\exp\left\lbrack
ig\int_{t_1}^{t_2}d{t'}
A_0^\alpha({\bf x},t')T_\alpha \right\rbrack \,\,\,,
}
is a path ordered product along a temporal line
with $T_\alpha$ the usual Hermitian colour matrices.
One formally gets in this way ($t>0$):
\eqn\PH{ P(t)\ \equiinf\
P_{stat}(t) \equiv \mbt \ \langle\ { Tr}
\left \{ {1- \gamma_{0} \over 2}\  \gamma_{0}\gamma _{5}
\ q({\bf 0},t)\  \bar q({\bf 0},0)\  \gamma_{0}\gamma _{5}\  \CP_{\bf
0}{t\choose 0} \right\} \rangle \,\,.
}
Except for the {\it c}-number factor $\mbt$, the operator whose vacuum
expectation value is taken in \PH\ depends only on light quarks and
gauge fields. It is a gauge-invariant quantity, as expected.
On the lattice the Green function corresponding to the bracket in eq. \PH\ is:
\eqn\PLAT{
 P_{L}( \tau ) = \langle\ { Tr} \biggl({ {1 + \gamma _0} \over 2}
S_{L}^{q}({\bf 0}, \tau ; {\bf 0}, 0)   U_{0}(1)
U_{0}(2) ... U_{0}(\tau - 1) \biggr)\rangle}
where $\tau = {t / a}$,  $U_\mu(x)$  is the link matrix pointing from
the site $x$ in the direction $\mu$, and $S_L^{q}$ is the light-quark lattice
propagator. The na\"\i ve $a \rightarrow 0$ limit transforms \PLAT\ into \PH\
and thus a Monte Carlo  estimate of \PLAT\
 would give an estimate of \PT, i.e.  of
$f_{_B}$.
However,  the breakdown of the static
approximation for the propagator \SB\  at high
frequencies (larger than $m_b$) and, as usual, the  lattice granularity induce
renormalization effects, so that we expect a relation of the type:
\eqn\Zdef{P(t) \equiinfa\
 Z P_{MC}(\tau)\  e^{-E_B a \tau}
, }
where $P_{MC}$\ is the numerical value of $P_L$ [eq. \PLAT ]\ as measured in
Monte Carlo simulations;
$E_B$ is introduced to match the time dependence of the two sides
of \Zdef; its role and significance will be discussed in more detail later.
 The renormalization constant $Z$ is computed by comparing in perturbation
theory  the corresponding continuum and lattice quantities\foot{For simplicity,
we will use a massless light quark, since none of our
conclusions depends on this choice \LIN, \EICH. The Feynman gauge is used
everywhere. Continuum renormalizations have been performed, as in ref. \LIN,
in the $\overline {MS}$ scheme supplemented
by an on-shell mass renormalization.}.

To this end two methods have been proposed.
The first one (BLP method) \LIN\ consists in a direct perturbative
comparison of the correlation functions \PT\ and \PLAT.
To order $g^2$ this amounts to evaluating
the two-loop diagrams  of Fig.~\fig\new{The four contributions of order
$g^2$ to $P(t)$, $P_{stat}(t)$ and $ P_{L}( \tau )$  as defined in eqs \PT,
\PH\ and \PLAT\ respectively (in order to avoid
redundancy, we use the same graphic representation for the three
cases).}\foot{In order to
avoid redundancy, we use the same graphic representation for the three cases we
shall encounter. According to the situation, the heavy line must be interpreted
as a continuum quark propagator \PT, a lattice static quark source \PLAT\ or a
continuum static quark source \PH. The context should make it clear which
interpretation is to be used in each case.}.
It will also be convenient to compute \PH\ as an intermediate step, which
involves considering the graphs of Fig.~1 in the case of
a static continuum quark source.
The contributions to $Z$ coming from the diagrams of Fig.~1
labelled by $b$, $c$ and $d$
will be called $Z_{light}$, $Z_{vertex}$\ and $Z_{heavy}$\ respectively.
To the order we work we have
\eqn\ZZZ{
Z= Z_{light}Z_{vertex}Z_{heavy}\ \ .}

The EH method of ref. \EICH\ relies on the remark that the dynamics
of a static (colour triplet) quark source
can be represented by the effective Euclidian Lagrangian:
\eqn\LAGC{
{\CL}_{EH} ={\rm B}^\dagger \left( \partial_0 + i g A_0\right) {\rm B}
\,\,\,,
}
where the B (${\rm B}^\dagger$) field annihilates (creates) a static quark.
It should be noted that a fixed four-momentum $(m_b,{\bf
0})$ has been removed from the momentum of the heavy quark. A
completely independent static field should be introduced to describe
processes involving
heavy antiquarks. The B field is a two-component spinor that describes spin-up
and spin-down static $b$ quarks.

The free propagator in position space is given by:
\eqn\PROPEFF{
 S_{EH}(x,y)=\langle {\rm B}(x) {\rm B}^\dagger(y)\rangle ={\theta(x_0-y_0)\
\delta({\bf x}-{\bf y})}\,
\,\,\,,
}
and in momentum space by:
\eqn\PROC{
 S_{EH}(p)={{i\over p_0+i\epsilon}}\,
\,\,.
}
Equation \LAGC\ can be formally  obtained starting from the Dirac action
of a particle of  mass $m_b$ and integrating out the ``small components''
 of the Dirac field in the limit $m_b \to \infty$.

The lattice-discretized version of the Lagrangian \LAGC\ can be
chosen in Euclidean space to be \EICH:
\def\zerohat{\hat 0}
\eqn\LAGLAT{
{\CL}_{EH}^L =
{\rm B}^\dagger(n)
\left({\rm  B}(n)-\left[U_0(n{-}\zerohat)\right]^\dagger
{\rm B}(n{-}\zerohat)\right) \, \, \,,
}
with $n\ \equiv\ ({\bf n},\tau)$,
leading to the free propagator for the lattice static source:
\eqn\PROLAT{
 S_{EH}^{L}(p)={{1 \over -{i\over a}(e^{i p_0 a}-1)+i\epsilon}}\,\,
\,\,.
}
In an external gauge field the exact static-quark propagator
coming from the action \LAGC\ [resp. \LAGLAT] leads indeed
to eq. \PH\ [resp. \PLAT].

In the presence of a ``local mass'' term
$ Ma {\rm B}^\dagger(n) {\rm B}(n)$\ the free propagator
coming from \LAGLAT\ becomes
\eqn\tplusun{<{\rm B}(n) {\rm B}^\dagger(0)> = e^{-Ma(\tau+1)}\theta (\tau )
\delta_{{\bf n},{\bf 0}}\ \ .
}

The main reason to introduce an effective action for the heavy quark
is  to reduce the determination of the renormalization coefficient $Z$\
 to the usual calculation
of vertex and self-energy corrections, \ie\   of the
diagrams  of Fig.~\fig\newbis{The graphs contributing to $P(t)$ in
an effective-action formulation.}. In this way, to order $g^2$, it
will be sufficient to compute one-loop graphs, whereas, in the BLP method,
one has to deal with two-loop calculations.

It should be noted however that  one has to face, in the effective-action
method, the problem of principles that the correlation function we are
interested in appears as a product of two operators at different times, but at
{\it the same space position}. Since the effective theory is not
Lorentz-covariant, the renormalization procedure is not  straightforward  and
it
is not obvious that complications, coming from an interplay between infrared
singular behaviour in time and ultraviolet divergences in space, will not be
present. In this work we shall prove the important fact that indeed such
complications do not arise in the one-loop calculation of $Z$. In spite of
that,
the two papers \EICH\ and \LIN\ give different results. The second purpose of
this paper is then to study in detail the nature of the normalization condition
that determines $Z$.

We will see that, as a matter of fact, the difference between the values
of $Z$ quoted by different authors is due to terms
 that originate, in the language
of ref. \EICH, from the freedom in the choice of the expression
for the effective heavy-quark action on the
lattice, while in the language of ref. \LIN\ they can be seen to come out
from an intrinsic (multiplicative) ambiguity in the form of the normalization
condition that determines $Z$ (see also \ref\MAIA{L. Maiani, G.
Martinelli and C.T. Sachrajda, Nucl. Phys. {\bf B 386} (1991) 281}).
Although these terms vanish as
$a \rightarrow 0$, in the present-day
simulations they are numerically not negligible. We
shall prove, however, that problems related to this
ambiguity can be eliminated, to leading order in $a$, by a consistent use of an
$O(a)$-improved action  for the light quarks \ref\ROS{G. Heatlie, C.T.
Sachrajda, G. Martinelli,  C. Pittori and G.C. Rossi, Nucl. Phys. {\bf B 352}
(1991) 266.},  together with the appropriate form of the
normalization condition.

The paper is organized as follows. In section~2 we discuss in detail
the nature of the normalization condition that allows us to relate, for
large $t$, the physical Green function $P(t)$ to the Monte Carlo measured
quantity $P_{MC}(\tau)$. The detailed comparison of the approaches
of refs. \LIN\ and \EICH\ is presented in section~3, where the controversial
calculation of the self-energy of the heavy-quark line is discussed.
We show that numerically the difference between the two
approaches lies in the way the ``wave-function'' renormalization
constant of the heavy line is extracted from the relevant Feynman diagrams.
In particular we notice that in the scheme of ref. \EICH\ the result
depends upon the way the heavy-quark action is discretized on the lattice.
section~4 addresses the problem  of the potentially dangerous
interplay between infrared (in time) and ultraviolet (in space)
singularities in the vertex diagram.
We rigorously prove that, to the order at which we work, this
phenomenon does not occur, and
the two methods lead to the same result for $Z_{vertex}$. Along the way we
correct an error present in ref. \LIN. In section 5 we give
the final numerical values of $Z$.

\newsec{The normalization condition}

The value of $Z$ in \Zdef\ can be obtained from the comparison of the large
$t$ behaviour of the non-perturbative ratio
\eqn\RNP{R_{NP}=
{<  J_{qb}^\dagger({\bf 0},0)\ J_{qb}({\bf x},t)>\vert _{phys}
\over P_{MC}(\tau )\ }\  \equiinfa\
{ {f_{_B}^2M_B\over 2} { e^{-M_Bt}}\over C_{MC} e^{-Bt} }
}
to the corresponding perturbative one
\eqn\RPT{R_{PT}=
{<  J_{qb}^\dagger({\bf 0},0)\ J_{qb}({\bf x},t)>\vert _{PT}
\over  P_{L}(t)\vert _{PT} }\ \ .}

In eq. \RNP\ $B$ is a mass parameter fitted from the (expected) time
exponential behaviour of $P_{MC}(\tau )$:
\eqn\PLMC{P_{MC}(\tau)\  \equiinfa\  C_{MC} e^{-Bt}\ \ .}
To any order in perturbation theory the behaviour of the r.h.s. of eq. \RPT\
is, actually, a polynomial in $t$. However in the large $t$ limit,
this polynomial is the $g^{2}$
expansion of a single exponential. In the following, for brevity, we will
simply refer to it as an exponential.

With this in mind, one can evaluate Z through the following two-step
procedure:
\item{1)} given the quantity, indicated as $R_{NP}$ in \RNP, one computes
in perturbation theory exactly the corresponding ratio of Green
functions, $R_{PT}$;
\item{2)} once the large $t$ behaviours of
$R_{NP}$ and $R_{PT}$ have been removed, the two remaining constant quantities
are declared to be  equal in the limit of vanishing lattice spacing.
In formula \RPT, having defined $M_{PT}$
so as to perturbatively remove the exponential $t$ behaviour of $R_{PT}$, i.e.
\eqn\DEFMPT{\lim_{a\tau = t \to \infty} e^{M_{PT}t} R_{PT}(t) ={\rm const} \ ,}
the normalization condition can be expressed as:
\eqn\NORMC{{{f_{_B}^2M_B\over 2}\over C_{MC}} =
   \lim_{a \to 0 }\ \lim_{a\tau=t \to \infty}
e^{M_{PT}t} R_{PT}(t) \ \equiv\ Z\ \ .}
For future reference we write:
\eqn\MB{M_{PT} = \delta_{PT} - \sigma\ \ ,}
\noindent{where, by definition, $\sigma$ is the perturbative part of $B$:}
\eqn\B {B = \sigma + \delta_{NP}\ \ .}

We note that, although $\sigma$ is, as we shall see, linearly divergent as
$a \to 0$, the non-perturbative quantity $\delta_{NP}$ is regular, i.e.
\eqn\Bsig {\lim_{a \to \ 0} B - \sigma =
\lim_{a \to \ 0} \delta_{NP}\ =\ {\rm finite}}

Two observations are in order here. The first one is that the
normalization condition \NORMC\ has to be imposed in the continuum
limit, i.e. among finite quantities
in the limit $a \to 0$ and $a\tau =t$ fixed. The second, and more
important one, is that eqs.~\DEFMPT\ and \NORMC\ can fix $Z$ only up to
a multiplicative factor (going to 1 as $a \to 0$). In fact the
condition \NORMC\ does not forbid multiplying $R_{NP}$ by a time-independent
constant,  provided we similarly multiply $R_{PT}$ by the corresponding
perturbative factor, as prescribed by the normalization procedure described
above.
 For instance we can
equally well replace eq. \NORMC\ by: 
\eqn\NORMCP{\lim_{a \to 0}{{f_{_B}^2M_B\over 2}\over {e^{-xaB} C_{MC}}} =
\lim_{a \to 0 }\ \lim_{a\tau=t \to \infty}
{{e^{M_{PT}t} R_{PT}(t)} \over e^{-xa\sigma}}\ \ ,}
where $x$ is any (small) real number. Equation \NORMCP , for $x = 1$, is
essentially the normalization condition imposed in ref.~\EICH\ (see section~3),
while for $x = 0$ we get the prescription of ref.~\LIN.

The exponential $\exp(-xa\sigma)$ in the r.h.s. of eq. \NORMCP\ is
introduced to compensate for the corresponding factor, $\exp(-xaB)$,
present in the l.h.s. This is in agreement with our prescription and it
is necessary because, as we have said,
$B$ is (perturbatively) linearly divergent, so $aB$ does not vanish as
$a \to 0$, while $a(B - \sigma)$ does.

Using eq. \NORMC\ or \NORMCP, one can extract $f_{_B}$ from Monte Carlo
data. The reason why in practice the two choices, $x=0$ and $x=1$, lead
to different
values for $f_{_B}$ is then obvious. Since $P_{MC}$ and, consequently, $B$
is only known at finite values of $a$, the two exponentials
 in eq.~\NORMCP\ do not compensate
exactly. If $\sigma$ is  taken to one-loop in
perturbation theory, one has: %
\eqn \EBS {\exp[xa(B - \sigma)] = 1 + x [ O(a) + O(g^4)] \ \ne\ 1\ \ .}
As a consequence, the actual value of $f_{_B}$, extracted from Monte Carlo
data, depends on the detailed form of the normalization condition. From
eq.~\NORMCP\ one gets:
\eqn \FBX {(f_{_B})^2_x = (f_{_B})^2_{x=0} \exp [xa(B -\sigma)]}
where, using the definition of $Z$ given by eq.~\NORMC, one has
\eqn \FBdef {{M_B \over 2}\ (f_{_B})^2_{x=0} = Z\ C_{MC}\ \ . }
For $x=1$, eq.~\FBX \ becomes
\eqn \EHBPL {(f_{_B})_{_{EH}} = (f_{_B})_{_{BLP}}\  \exp\left[{a(B - \sigma)
\over 2}\right]} %
which gives the relation between the value of $f_{_B}$
extracted from Monte Carlo data following the approach of ref.~\EICH, called
here $(f_{_B})_{_{EH}}$, and $(f_B)_{_{BLP}}$, the same quantity as evaluated
within the approach  of ref.~\LIN .

Strictly speaking, there is no compelling reason to prefer $x=0$
to $x=1$, or, for that matter, to any other value of $x$. We shall
see, in fact, in subsection~3.4 that exactly the same line of arguments,
which ``naturally'' led to the choice $x=1$ in ref.~\EICH, can
also be used to arrive ``naturally'' at the value $x=-1$.

The question is then whether there is any criterion to fix
this problem univocally. It is not difficult to see that this can
be done if we decide to improve the light-quark Wilson action
\ROS\ employed in the Monte Carlo
simulation of $P_{MC}$ and in the corresponding perturbative
calculations of $Z$ \ref \ARCA {A. Borrelli
and C. Pittori, Department of Physics, University of Rome
 ``La Sapienza'' preprint n.864 (February 1992).}, \ref\HEHI{O.F.
Hernandez and B.R. Hill, Los Angeles preprint UCLA/92/TEP/9 (March 1992).}.

In this case, in fact, since eq.~\Bsig\ maintains its validity, the only
consistent way of eliminating {\bf all} $O(a)$ terms from the lattice
evaluation of $f_{_B}$  is to take $x=0$, that is to use the normalization
condition \NORMC. If we do so, we remain with only $O(ag^2) + O(g^4)$
corrections.

\newsec{Heavy-quark renormalization}

In this section we want to compare in detail the approaches
of refs.~\LIN\ and \EICH\ and to clarify the origin of the discrepancy between
the two results by discussing the calculation of the
graphs of Figs.~1 (d) and 2 (c) which give rise to the controversial
contribution, $Z_{heavy}$, to $Z$.

\newssec{BLP evaluation of $Z_{heavy}$}

With the BLP method the graphs of  Fig.~1 (d) have been computed  in
ref.~\LIN . In the Feynman gauge, their contributions to \PLAT, \PH\ and
\PT\ are respectively:
\def\intpipi{\int\limits_{-\pi}^{+\pi}{d^4{\rm k}\over (2\pi)^4}}
\eqn\PDLAT{
\eqalign{P_L^{d}(t)\ &=\ -{4\over 3}g^2\intpipi
{1-\cos(k_0t/a) \over 1-\cos(k_0) } {1 \over 2\sum\limits_\lambda
(1-\cos(k_\lambda))}\  P_L^{tree}(t)\cr
&\equiinf \ \left\{\ {g^2 \over 3\pi^2}\left[ \ln\left({ t \over 2a}\right)
+{\gamma_E \over 2}+{F_{0000}+F_{0001}\over 4}
+1 \right] - \sigma a \tau \right\} P^{tree}(t)\  \cr}
}
\eqn\PDH{
P_{stat}^d(t)\ =\ \mbt\ {g^2 \over 3 \pi^2}\left[ \ln \left( \mu t \over
2\right)+1+\gamma_E\right] \ P^{tree}(t)\ ,\qquad \qquad
\qquad \qquad
}
\eqn\PDC{
P^d(t)\ =\ \mbt\ {g^2 \over 3 \pi^2}\left[ \ln \left({m_b t
\over 2 }\right) + {1\over 2}\ln \left(m_b\over\mu \right)+
\gamma_E \right]\
P^{tree}(t)\ .\qquad \qquad} %
where, on the lattice:
\eqn\Ptreelatt{\eqalign{ P_L^{tree}(t)\ &=\ \,
\intpipi \,{1\over2}\cr &\qquad  \times  Tr\left [
(1+\gamma_0) { -i \sum_\mu \gamma_\mu \sin p_\mu
+ma+r\sum_\nu(1-\cos p_\nu)\over \sum_\mu \sin^2 p_\mu
+\left( ma+r\sum_\nu(1-\cos p_\nu)\right)^2}
\right ]\cr
&\equiinf\,\, \,P^{tree}(t)\ \,\,. }}

\noindent and in the continuum

\seqnn\Ptreecont{P^{tree}(t)\ =\ {3\over\pi^2 t^3}\ \ .  }

In \PDLAT, $F_{0000}$ and $F_{0001}$ are the numerical constants defined in
ref.
\ref\FF{A. Gonzales Arroyo and C.P. Korthals Altes, Nucl. Phys. {\bf B 205}
(1982) 46.}: $F_{0000} \simeq 4.369$ and $F_{0001} \simeq 1.311$; $\gamma_E
= 0.5772 $ is the Euler constant,  and
\eqn\DELTAMB{
\eqalign{
\sigma=&{g^2 \over 24\pi^3}{1\over a}
\int\limits_{-\pi}^{+\pi}d^3k{1\over
\sum\limits_j(1-\cos k_j)}\cr
\simeq &{g^2 \over 4\pi^2}{1\over a} \cdot  6.65
}}
is the linearly-divergent one-loop  radiative correction to the energy
of the static source.
In the following we will often write $1-a\sigma
= e^{-a\sigma } +O(g^4)$.

 From these results we get for the $Z_{heavy}$ contribution
 to the total normalization  constant $Z$, defined
in \ZZZ\ and \NORMC:
\eqn\ZDHMOINSZDL{\eqalign{
 Z_{heavy}\ -\ 1\ &\equiv\  \lim_{t \to \infty}\ e^{M_d t}
\ { P^{tree}(t)\ +\ P^d(t) \over \bigg( P^{tree}_L(t)\ +\ P^d_L(t)\bigg)}
\ -\ 1\cr &= {g^2\over 3\pi^2}\left\{ \ln(m_b a)
+ {1\over 2}\ln \left(m_b\over\mu \right) +{\gamma_E \over 2}
-{{F_{0000}+F_{0001}}\over 4}-1\right\} \cr
&\simeq {g^2\over 4\pi^2}\left\{ {4\over 3}\ln(m_b a) +
{2\over 3}\ln \left(m_b\over\mu \right) -2.84 \right\}
 \ \ , }}
where $e^{M_dt}$ is the appropriate factor necessary to compensate for the
overall time exponential behaviour in the r.h.s. of the first line of
 eq.~\ZDHMOINSZDL. In perturbation theory the multiplication by this factor
simply amounts to subtracting out from the computed amplitudes all terms
proportional to $t$.

\newssec{EH evaluation of $Z_{heavy}$}

With the second method \EICH, the self-energy of the heavy quark,
 $\Sigma(p_0)_{_{EH}}$, is
computed directly with the Feynman rules deduced from the effective action
\LAGLAT, obtaining:
\eqn\SIGMAEICH{
\eqalign{\Sigma(p_0)_{_{EH}}\ = \ &{4\over 3}g^2{1\over a}\intpipi \
e^{i(k_0+2ap_0)} {1 \over 4\sum\limits_\mu \sin^2({k_\mu \over 2})}\ {i \over
e^{i(k_0+ap_0)}-1-\epsilon}\cr
&-{2\over 3}g^2 {1\over a}\ e^{iap_0}\intpipi{1\over 4\sum\limits_\mu
\sin^2({k_\mu \over 2})}\ \ \cdot }
}

It is crucial to remark that if we insert
$\Sigma(p_0)_{_{EH}}$ into the external quark loop, we recover exactly
the integral giving
$P_L^d$ in eq.~\PDLAT. So the two methods should agree. In ref.~\EICH,
the wave function renormalization was computed with the standard formula:
\eqn\RENORM{ Z_2-1 \ =\ -i{\partial \Sigma
 \over \partial p_0}\bigm|_{p_0=0}\ \ .
  }
It is immediate to check that the radiative correction to the mass is the
same as in \DELTAMB, \ie
\eqn\DMCOMP{
\sigma \ =\ \Sigma(p_0=0)\ \ .}

Using eq.~\RENORM, the result for the ratio of the wave function
renormalization constant of the  continuum dynamical $b$-quark [eq.~\PT ]
to that of the lattice static  $B$\ field in \LAGLAT\   is:
\eqn\RAPSIG{{1-i{\partial \Sigma_{cont} \over \partial p_0} \over
1-i{\partial \Sigma_{_{EH}} \over \partial p_0}} \Bigg|_{p_0=0}\
=\ Z_{heavy}\ {(1-a\sigma)}\simeq Z_{heavy}\ e^{-a\sigma }\ \ ,}
with $Z_{heavy}$ given by eq.~\ZDHMOINSZDL.

\newssec{Comparison of the two approaches}

To compare the results for $Z$ obtained in refs.~\LIN\ and \EICH,
one has to include in the calculation the values of $Z_{light}$
and $Z_{vertex}$.
As for  $Z_{light}$, it is easy  to see that the analog of the ratios in
 eqs.~\ZDHMOINSZDL\ and \RAPSIG\  for the light-quark
self-energy diagrams, explicitly given in subsection~5.1 [Figs.~1 (b) and
2 (a)], yield the same results,
both with the method of ref.~\EICH\ and with that of ref.~\LIN.
The reason is that there is no ambiguity of the type we have discussed above
for the heavy $b$ quark. In fact, on the one hand the way the
mass-counterterm for the light quark is introduced is univocally fixed
by the form of the Wilson action that is being used, and, on the other hand,
both in  the Monte Carlo simulations and in the perturbative calculations the
same expression for the light-quark action with the same mass renormalization
condition  (vanishing pion and quark masses respectively) is employed.
Concerning the contribution of the vertex diagrams [Figs.~1 (c) and
2 (b)], we will show  in section~4 that the results of refs.~\LIN\ and
\EICH\ for $Z_{vertex}$ also coincide.

It then follows that the relation between $Z$\ in \Zdef, as computed in
ref.~\LIN\ and the current renormalization constant $Z_J$ introduced
in \EICH\ is
\eqn\ZJZ{
Z_J^{-2}\ =\ Z\ e^{-a\sigma}.}
Now we turn to the central issue: given a Monte Carlo data set, what
is the prediction for the physical \FB, and do the two methods agree?

Fitting the Monte Carlo data for large $t$ through formula \PLMC,
in ref.~\LIN\ one arrives at the result
\eqn\FBBLP { {f_{_B}^2 M_{B} \over 2 }|_{_{BLP}}\ =\ Z C_{MC} \ \ .}

In ref.~\EICH , the Monte Carlo data are fitted instead, in view of
eq.~\tplusun, according to
\eqn\MCEUX{
P_{MC}  \equiinf \ A\ e^{-B(t+a)} \ \ .}
Comparing \PLMC\ and \MCEUX\ one has
\eqn\CMC {C_{MC}\ =\ A e^{-Ba}\ \ .}
For the physical correlation function (eq. (26) in \EICH) one obtains:
\eqn\TWOGET{ {f_{_B}^2 M_B \over 2}|_{_{EH}}\ e^{-M_B
 a \tau}\ =\ Z_J^{-2}\ e^{-m_b a\tau}\ A\
e^{-Ba(\tau+1)}e^{-\Delta a(\tau+1)}\ \ .}
For consistency of our notation we have rewritten the factor
$1/(1+a\delta m/Z)$\  of ref.~\EICH\ as
$e^{-a\Delta}$. By matching the large $t$ behaviour of \TWOGET, one gets:
\eqn\RESDEL{\Delta = M_B-m_b-B}
and
\eqn\RESEH{
{f_{_B}^2 M_B \over 2}|_{_{EH}}\ = Z_J^{-2}Ae^{-Ba}e^{-\Delta a} =
Z_J^{-2}Ae^{-(M_B-m_b)a}\ \ .}

At this point the authors of ref.~\EICH\ identify $M_B - m_b$ as the
 $B$-meson binding energy. They estimate $\Delta$ in the limit $a \to \ 0$,
by arguing that $M_B - m_b$ must converge to a constant in the continuum
limit, i.e. $\displaystyle{\lim_{a \to 0}}
 a(M_B-m_b)=0$, so they end up with:
\eqn\RESEHF{{f_{_B}^2 M_B \over 2}|_{_{EH}}\
=\ Z_J^{-2}\ A\ \ .}

Remembering the relations \ZJZ\ and \CMC, the comparison of \FBBLP\ and
\RESEHF\ gives:
\eqn\COMP{f_{_B} |_{_{EH}}\ =\ f_{_B} |_{_{BLP}} \ \times e^{{a(B-\sigma) \over
2}}\ \ .}
This answers our central question: {\it given a set of Monte Carlo data
the number extracted for the physical $f_{_B}$ according to \EICH\
 [eq.~\RESEHF] will
be  that of \LIN\ [eq.~\FBBLP]  multiplied by
$\exp[{{a(B-\sigma) \over 2}}]$}.

Since $a(B-\sigma)=a(-\Delta-\sigma+M_B-m_b)$\
is
 $O(g^4)+O(ag^0)$, it follows
that {\it both results agree in principle in the approximation in which we
are working}. However,
taking a typical Monte Carlo slope $aB=0.7$\ ($aB=0.5$) for $\beta=6$
($\beta=6.4$), eq.~\RESEHF\ gives
a value for $f_{_B}|_{_{EH}}$\ that is larger by a factor $\sim 1.3$\ ($\sim
1.2$) than
 $f_{_B}|_{_{BLP}}$\ for the same Monte Carlo data set: the discrepancy has an
obvious practical relevance and its origin should be fully understood.

\newssec{Further remarks}

Let us  insist once more on the content and the origin of the factor in \COMP.
First, it has nothing to do with the computational method, namely the use of
either  direct computation of the correlation function or effective
Lagrangians. Second, there is no discussion on the validity of the
starting equation \Zdef. Thirdly there is no discrepancy in the mathematics.

The point is that, in ref.~\LIN, $Z$ is computed (to second order of
perturbation theory) by using the normalization condition \NORMC, thus
directly obtaining eq.~\FBBLP. In this way, one must simply assume that an
$E_B$ exists, which matches the time dependence of both sides of eq.~\Zdef,
without the drawback of having to give a problematic estimate of a linearly
divergent quantity. As we have discussed in section~2, however, an apparently
irrelevant modification of the normalization condition, such as the one
made in eq.~\NORMCP, leads to a non-negligible numerical ambiguity in
the  estimated value of $f_{_B}^2$ given by the factor \EBS.

The approach of ref.~\EICH , instead, relies on the use of an effective
Lagrangian for the heavy quark, both in the way the large {\it t} dependence
of the perturbative and the Monte Carlo expressions of $P_L$ is parametrized
and in the way the wave-function renormalization of the heavy-quark line is
computed. The authors of ref.~\EICH\ were, in fact, led by the form \tplusun\
of their free propagator to the parametrization \MCEUX; they were also driven
by formula \RENORM, which looks so familiar, to choose $Z_2$ as the
quantity to be computed perturbatively.

The key observation at this point is that the effective action is merely an
intermediate step used to compute the relative normalization between the
physical observable and the quantity measured on the lattice. Neither quantity
has anything to do with any effective action and the final result should not
depend on it. However, it is easy to produce a variety of effective lattice
actions whose continuum limit is \LAGC\ and which, by the use of the
corresponding matching condition and of formula \RENORM, lead to different
values of $f_{_B}$ for a given set of Monte Carlo data.

For instance instead of \LAGLAT, one could as well choose for the discretized
version of \LAGC\ the Lagrangian:

\eqn\LEHP{
\CL^{\prime}_{EH} \ =\ {\rm B}^\dagger(n) \left( U_0(n+ \zerohat) {\rm B}(n
+ \zerohat) \ -\ {\rm B}(n)\right)\ \ .}
If a local mass term  $Ma {\rm B}^\dagger(n){\rm B}(n)$\ is added, one
gets:
\eqn\PROPP{<{\rm B}(n) {\rm B}^\dagger(0)> = e^{-Ma(\tau-1)}\theta (\tau )
\delta_{{\bf n},{\bf 0}}\ \ .}

Notice the change $(\tau +1) \rightarrow (\tau-1)$ in eq.~\PROPP\ with respect
to eq.~\tplusun. This  change propagates step by step from eq.~\MCEUX, leading
now to:
\eqn\FBP{
f^{\prime}_{_B}|_{_{EH}}\ =\ f_{_B}|_{_{BLP}} \times e^{-{a(B-\sigma) \over
2}}}
instead of \COMP.

Another interesting choice  is to use an $O(a)$-improved  effective
Lagrangian for the heavy quark, i.e. to take:
\eqn\LAGIMP{\eqalign{
{\CL}_{imp}= &\
{\rm B}^\dagger(n)
\Biggl({3\over2} {\rm B}(n) - 2 U_0(n{-}\zerohat)^\dagger {\rm
B}(n{-}\zerohat)\cr & \ +{1\over 2}U_0(n{-}\zerohat)^\dagger
U_0(n{-}2\zerohat)^\dagger {\rm B}(n{-}2\zerohat)  \Biggr)
\, .}
}
One checks that, if the static field is integrated out, the Lagrangian
\LAGIMP\ leads exactly to the same correlator \PLAT\ as the Lagrangians
\LAGLAT\ or \LEHP, up to exponentially small terms, vanishing
as $t \to \infty$.\par
In view of this argument, the authors of \ARCA\ and \HEHI\ claim correctly
that there is no need to use an improved effective action for the heavy
quark, if one uses the normalization condition \NORMC . However,
if one insists on using eq.~\RENORM\  to define
$Z_{heavy}$ the precise form of the effective action does matter. In
fact, the free propagator obtained from \LAGIMP\ is:
\eqn\PROIMP{
 S_{imp}(p)={{1 \over -{i\over a}( 2 e^{i  p_0 a}-
{1\over 2} e^{i 2 p_0 a}-{3\over
2})+i\epsilon}}\,\, \,\,.
}
It coincides with the continuum one [eq.~\PROC] up to $O(a^2)$, while
\PROLAT\ differs already at $O(a)$. We have computed $\Sigma_{imp}(p)$, the
self-energy of the heavy line, with the ``improved'' effective Lagrangian
\LAGIMP\ and derived the
mass and wave function renormalization, by using formulae \RENORM\ and
\DMCOMP;  $\Sigma_{imp}(0)$ is identical to \DELTAMB, but the noteworthy fact
is that, instead of eq.~\RAPSIG, one obtains
\vskip .15in
\eqn\RAPIMP{{1-i{\partial \Sigma_{cont} \over \partial p_0}
\over
1-i{\partial \Sigma_{_{imp}} \over \partial p_0}}\Bigg|_{p_0=0}\
=\ Z_{heavy}\ \ . }
\vskip .15in
Following the steps of ref.~\EICH , we end up this time with a result identical
to \FBBLP\ (the same as in ref.~\LIN), that is:
\eqn\FFIMP{\left( f_{_B}\right)_{imp} |_{_{EH}}\ =\ \left( f_{_B}\right)_{x=0}
\ =\  \ f_{_B} |_{_{BLP}}\ \ .}
Thus a determination of $Z$ that uses formula \RENORM\ can give the
same result as \NORMC, if one uses the Lagrangian \LAGIMP.

 From all this analysis, we can conclude that the
use of the effective action approach by itself is not sufficient to lead to an
unambiguous (in the sense we are discussing here) determination of
$f_{_B}$\ from Monte Carlo simulations.

We have instead shown in section~2 that this can be achieved to $O(a)$ only if
one consistently uses:
\item{i) }an $O(a)$ improved action for the {\bf light} quarks.
 \item{ii) }the normalization condition \NORMC, in which the
matching does not involve unknown $O(a)$ terms.

\newsec{Vertex renormalization}

In this section we discuss the evaluation of the contribution to $Z$ of the
vertex diagrams [Fig.~1 (c)]  and we rigorously prove that the effective
Lagrangian method leads to the same result as the full-fledged two-loop
calculation of ref.~\LIN. That is to say, there is no dangerous
interplay between IR and UV divergences.
The direct evaluation \LIN\  of the two-loop graph of Fig.~1 (c) gives:

\eqn\PCLAT{P_{L}^c(t)\ =\ {4\over 3}g^2{1\over t^3}\left(\CI_1^{latt}(\tau) \
+\ \CI_2^{latt}(\tau) \right)\ \ ,} %
with

\eqn\IUNETDEUX{\eqalign{
\eqalign{
      {\cal I}_1^{latt}=
       3\int\limits_{-\pi}^{+\pi}{dp_0 \over 2 \pi}
                 &\int\limits_{-{\pi \tau }}^{+{\pi \tau}}
{d^3 {\bf p}\over (2\pi)^3}     e^{ip_0\tau }
                    {
1-e^{ip_0}+\Sigma({{\bf p}\over \tau} ) \over
                       2 \Delta_2(p_0,{{\bf p}\over \tau} )
                                } \times
                                                          \cr
        e^{-ip_0}&\int\limits_{-\pi}^{+\pi}
          {d^4 {\rm k} \over  (2\pi)^4}
\ {e^{ik_0/2} \over e^{ik_0} -1}   {1\over \Delta_1 (k)}
 \ \ \bigg\{
           \ e^{-ik_0/2} {
        1-e^{i(p_0+k_0)}+\Sigma({{\bf p}\over \tau} +{\bf k})
 \over      \Delta_2(p_0+k_0,{{\bf p}\over \tau} +{\bf k})
                               }\cr
 &\qquad\qquad\qquad -\qquad
 \  e^{ik_0/2} {
               1-e^{i(p_0-k_0)}+\Sigma({{\bf p}\over \tau} -{\bf k})
 \over              \Delta_2(p_0-k_0,{{\bf p}\over \tau} -{\bf k})
                        }
   \ \ \bigg\} \cr
}
\cr
&  \cr
&  \cr
&  \cr
\eqalign{ {\cal I}_2^{latt}=3\sum_j
\int\limits_{-\pi}^{+\pi}{dp_0 \over 2 \pi}
&\int\limits_{-{\pi \tau}}^{+{\pi \tau}}{d^3
{\bf p}
       \over (2\pi)^3} e^{ip_0\tau}
{-i \sin({p_j\over \tau}) \over 2 \Delta_2(p_0,{{\bf
p}\over \tau})} \times
\cr  e^{-ip_0}&\int\limits_{-\pi}^{+\pi} {d^4 {\rm
k} \over  (2\pi)^4}\ {e^{ik_0/2} \over e^{ik_0} -1}{1\over
\Delta_1 (k)} \ \ \bigg\{
 \ e^{-ik_0/2} {-i\sin({p_j\over \tau}+k_j)
 \over \Delta_2(p_0+k_0,{{\bf p}\over \tau}+{\bf k})}\cr
 &\qquad\qquad\qquad -
\qquad \  e^{ik_0/2} {-i\sin({p_j\over \tau}-k_j)
\over \Delta_2(p_0-k_0,{{\bf p}\over \tau}-{\bf k})}  \ \
   \bigg\} \cr}
}}
\vskip 1cm

\noindent{where the $\Delta$'s come from the gluon and fermion
 lattice propagators:}
\eqn\DELTAS{\eqalign{
&\Delta_1(k)\ =\ {1\over 2}\sum\limits_\lambda
\left(1-\cos(k_\lambda)\right)\cr
&\Delta_2(p_0,{\bf p})\ =\ \sum\limits_\lambda \sin^2(p_\lambda)\ +\
  \left( \sum\limits_\lambda \left[1-\cos(p_\lambda)\right]\right)^2\ \ .\cr}}
As we said, we have chosen here to work with a massless light quark, since the
renormalization constants are independent of the mass \LIN, \EICH.
No mass has been given to the gluon since the quantities we are dealing with
are infrared-finite
 after the singularities among individual terms have been cancelled.

To determine $Z_{vertex}$, we have to study the
behaviour of $\CI_1^{latt}$ et $\CI_2^{latt}$ for large $t$, at fixed
$a$, \ie\ when
$\tau$ goes to $\infty$. The integrations over the temporal components,
$p_0$ and $k_0$, are performed exactly, by closing the contour in the complex
plane. Oscillating factors are turned into exponentially damped ones,
easier to handle. The large-$\tau$ behaviour of the resulting spatial integrals
is analysed by means of Lebesgue's lemma. We find (see appendix A for a
sketch of the proof):
\def\etauinf{\lower 0.12cm\hbox{$\widetilde{_{_{\tau\rightarrow \infty}}}$}}
\eqn\RESULTILAT{\eqalign{
&{\eqalign{\CI_1^{latt}(\tau)\ \etauinf\  {3\over \pi^2}&\intpipi
{e^{ik_0}-1-\sum_j (1-\cos(k_j))
  \over 2 \Delta_1(k) \Delta_2(k_0,{\bf k})}\ {1-e^{ik_0\tau} \over 1-e^{ik_0}}
\cr &+\ C_1 + o(\ln(\tau)/\tau)\ \ ,  \cr }}\cr&\cr
&\CI_2^{latt}(\tau)\ \etauinf \ C_2\ +\ o(\ln(\tau)/\tau) \ \ ,\cr }}
where $C_1$ and $C_2$ are some  ($\tau$-independent) continuum-like
integrals. A similar analysis
can be performed in the continuum for  $P_{stat}^c(t)$ [eq.~\PH], using
dimensional regularization $(d=4-\epsilon)$ to control its ultraviolet
behaviour, and gives:
\eqn\PCH{
P_{stat}^c(t)\ =\ {4\over 3}g^2{1\over t^3}\left(\CI_1^{stat}(\mu t) \
+\ \CI_2^{stat}(\mu t) \right)\ \ ,}
with
\def\equepso{\lower 0.12cm\hbox{$\widetilde{_{_{\epsilon\rightarrow
\infty}}}$}}
\def\simepso{_{ \ \sim \ \atop \epsilon \rightarrow 0 }}
\eqn\RESULTIH{\eqalign{
&\CI_1^{stat}(\mu t)\ \equepso \ {3\over
\pi^2}\mu^\epsilon \int\limits_{-\infty}^{+\infty} {d^{4-\epsilon}{\bf k} \over
(2\pi)^{4-\epsilon}} {e^{ik_0t} -1 \over { 1\over 2} (k^2)^2}\ +\ C_1\
+\ o(\epsilon)\cr
&\CI_2^{stat}(\mu t)\ \equepso \  C_2\ +  o(\epsilon)\quad \cdot \cr}}
\vskip .2cm
The constants $C_1$ and $C_2$ are the same as in eq.~\RESULTILAT. The
decomposition into the terms appearing in eqs.~\RESULTILAT\ and \RESULTIH\
were performed in such a way that  the dependence
upon the regularization scheme was isolated  in the first terms,
leaving $C_1$ and $C_2$ renormalization-scheme independent quantities.

Using the $\overline {MS}$ renormalization scheme in the continuum, we
find that the contribution to the ratio of \PH\ to \PLAT\ coming from this
graph (see appendix B) is:

 \eqn\ZCHMOINSZCL{ \eqalign{
\lim_{t \to \infty} e^{M_ct} &{P^{tree}(t) \ +
\ P^c_{stat}(t) \over P^{tree}_L(t)\
+\ P^c_L(t)}\ -1 \cr &\hskip 1.5cm \eqalign{&\eqalign{={g^2 \over
8\pi^2}{4\over 3} &\Bigg[  2\ln(\mu
a)+\gamma_E-F_{0000}-{1\over 2\pi} \int\limits_{-\pi}^{+\pi}d^3{\bf k} {1\over
\Delta_2(0,{\bf k})}\cr  -&{1\over
4\pi^2}\int\limits_{-\pi}^{+\pi}d^4{\bf k} \left( {\sum\limits_\lambda
\sin^4{k_\lambda \over 2} \over \Delta_1^2\Delta_2(k_0,{\bf k})}-{1\over 4
\Delta_2(k_0,{\bf k})}\right)
 \Bigg]}\cr
&\simeq - {g^2 \over 4\pi^2} \left\{ -{4\ln( \mu a)\over 3}+7.79\right\}\
\ ,}} }
where the factor $e^{M_ct}$ plays the same role as the similar exponential in
eq.~\ZDHMOINSZDL.\ The  result \ZCHMOINSZCL\ coincides  with the value
obtained from the evaluation of the one-loop
vertex graph of ref. \EICH. This indeed shows that there are at this
order no complications  due to the problem of the renormalizability of a
non-covariant effective theory. In ref.~\LIN\ this
computation  was actually performed  with a non-vanishing light-quark mass. The
results agree, apart from
the $1/2\pi \int d^3{\bf k} \Delta_2^{-1}(0,{\bf k})$ term, which is present
in eq.~\ZCHMOINSZCL\ but was erroneously forgotten in ref.~\LIN.

 \newsec{Results}
In this section we complete our discussion on the evaluation of $Z$,
by giving the expression of $Z_{light}$, and
we compare the values of \FB\ obtained with different normalization
conditions, with the number one would obtain by using  an $O(a)$-improved
light-fermion action [in which case it is the form \NORMC\ of the normalization
condition that, for consistency, has to be employed].

\newssec{Contribution of $Z_{light}$}

The last piece of the puzzle is the contribution to the
renormalization constant due to the graph \new\  (b) which is the usual
wave-function renormalization for a Wilson fermion \ref\LIGHT1{A. Gonzales
Arroyo, F.J. Yndurain and G. Martinelli, Phys. Lett. {\bf B 117} (1982)
437.}, \ref\LIGHT2{W. Hamber and C.M. Wu, Phys. Lett. {\bf B 133} (1983)
351.}, \ref\LIGHT3{C. Bernard, A. Soni and T. Draper, Phys. Rev. {\bf D 36}
(1987) 3224.}. Numerically one obtains:
\eqn\ZBHMOINSZBL{ e^{M_bt}
{P^{tree}(t) + P_{stat}^b(t) \over P^{tree}_L(t) + P_{L}^b(t)}-1\
 \equiinf
- {g^2 \over 4
\pi^2 } \left\{ {2\ln(\mu a) \over 3} + 4.29
\right\}\ \ , }
where, once again, $e^{M_bt}$ is the appropriate factor introduced to
make the ratio
in the l.h.s. go to a constant as $t \rightarrow \infty$.

To reach the final result, we need the contribution to the ratio
$P_{stat} / P_L$ coming from the d-type graphs in Fig.~1   [in analogy
with eqs. \ZCHMOINSZCL\ and \ZBHMOINSZBL]. One has:
\eqn\ZDH{
\eqalign{ e^{M'_dt}
{P^{tree}(t) + P_{stat}^d(t) \over  P^{tree}_L(t)\ +\
P^d_L(t)}-1\
& \equiinf
\ {g^2 \over 4
\pi^2 } \left\{ {4\ln(\mu a) \over 3} +{2\over3}\gamma_E
-{{F_{0000}+F_{0001}}\over 3}
\right\} \cr
&\simeq {g^2\over 4\pi^2}\left\{ {4\over 3}\ln(\mu a)
-1.51 \right\}} }

The relation between the two continuum correlation functions \PT\ and \PH\ is
already known \LIN, \EICH;  the result is the same in the two methods
and it has
also been checked in the temporal gauge ($A_0\ =\ 0$)\ref\AZERO{J.P. Leroy,
J. Micheli, G.C. Rossi and K. Yoshida, Z. Phys. {\bf C 48} (1990) 653.}.
It can be written:
\eqn\PHVERSUSPT{ e^{Mt} {P(t) \over P_{stat}(t)}\Bigg|_{PT} - 1 \equiinf
 -\ {g^2 \over 2 \pi^2}\ln\left[ {\mu e^{2/3} \over m_b}\right] \ \ . }

We insist once more on the fact that all the exponential factors in
eqs.~\ZDHMOINSZDL,\
\ZCHMOINSZCL,\ \ZBHMOINSZBL,\ \ZDH\ and \PHVERSUSPT\ are defined  so as to
insure  the constancy of the corresponding ratios of correlation functions as
$t \rightarrow \infty$.
The values of the mass coefficients are totally irrelevant for the purpose
of computing $Z$, because the only thing one actually has to do is to
neglect in the calculations all the terms that are proportional to $t$.
Combining eqs. \ZCHMOINSZCL,\ \ZBHMOINSZBL,\ \ZDH\ and \PHVERSUSPT,
we obtain the result:

\eqn\FINAL{
 e^{M_{PT}t} {P(t) \over P_L(a,t)} \Bigg|_{PT} = Z\ -\ 1 \ =\
{g^2 \over 4\pi^2} \left[ 2\ln\left( m_b a e^{-2/3}\right)\ -\
13.59\right]\ \ .
 }

\newssec{Numerical evaluation of $Z$}

The relation of the physical value of \FB\ to the measured lattice quantity
 $$\displaystyle f_{_B}^{latt}\  \equiv\ \left({2 C_{MC} \over
M_B}\right)^{1/2}$$
extracted from the behaviour of $P_{MC}(\tau)$  at large $t$
[eq.~\PLMC] is then, from eq.~\FINAL:
\eqn\FBFB{ f_{_B} = f_{_B}^{latt} \left\{ 1 + {g^2 \over 8\pi^2}
\left[ 2\ln \left(  a m_b e^{-2/3} \right) \ -\ 13.59 \right] \right\}\ \ .
}
Numerically, if we take  $m_b  = 4.5$\ GeV and $\beta  =  6/g^2 = 6.0, 6.2,
6.4$\ with $a^{-1}=2.3, 2.9, 3.7$\ GeV
respectively, which are typical values for the Monte Carlo
simulations used in $B$
mesons studies \ref\MONTE{M.B. Gavela, L. Maiani, S. Petrarca, G. Martinelli
and
O. P\`ene, Phys. Lett. {\bf B 206} (1988) 113; \hfill\break C.R. Allton et
al., Nucl. Phys. {\bf B} (Proc. Suppl.) {\bf 20} (1991) 504.},
we see that in all cases the
logarithmic correction is negligible and we obtain the final result:
\eqn\FIN{ f_{_B} = 0.83 \, f_{_B}^{latt}\ \ .}
Note that the prescription of ref.~\EICH\ [i.e. $x=1$ in eq.~\FBX ],
 would lead to much larger values for the renormalization constant $Z$.
As a consequence, one would obtain
  $f_{_B} \sim (0.97 \,\hbox{to}\, 1.07) f_{_B}^{latt}$
for typical Monte Carlo slopes of $0.5 a^{-1}$\ to $0.7 a^{-1}$. With
the choice $x=-1$ [eq.~\FBP ] the value of \FB would be lowered to
  $f_{_B} \sim (0.63 \,\hbox{to}\, 0.70) f_{_B}^{latt}$.

Recently the renormalization constants of operators involving heavy
quarks have been computed \ARCA\
 by using for the light quarks a nearest-neighbour $O(a)$ improved action.
In this theory \ROS\ $O(a)$ corrections are absent from on-shell hadronic
matrix elements. Using the normalization condition \NORMC, the net result
for $Z$ amounts to replacing the constant 13.59
in \FBFB\ by 10.08. Then to first order in perturbation
theory one gets:
\eqn\FINIM{f_{_B} = 0.87 \, f_{_B}^{latt-imp}\ \ ,}
where $f_{_B}
^{latt-imp}$\ is the Monte Carlo value measured in simulations
which employ the same improved action as is used in the corresponding
perturbative calculations. We stress again
that in this case the prescription \NORMC\ [$x=0$ in \EBS ] is
compulsory in order not to loose $O(a)$ improvement.

Two numerical calculations of $f_{_B}$\ with a static $b$-quark
have been performed in the literature
\ref\rall{C.R. Allton et al., Nucl. Phys. {\bf B 349} (1991) 598.},
 \ref\ralex{C. Alexandrou et al., Phys. Lett. {\bf B 256} (1991) 60.}  at
$\beta=6$,  using for the renormalization coefficient the value
$0.8$, essentially in agreement with the estimate \FIN\foot{It is argued in
refs.~\rall\ and \ralex\ that the value
$0.8$ is in essential agreement with the results of both \LIN\ and \EICH.
This statement
probably refers to the value $1.22$ found for $Z_{_J}$ in \EICH. However,
as we discussed in section~3, this apparent agreement leads in fact to
quite different values for \fb.}.
 A systematic comparison with the
results on meson decay constants for the case of propagating
quarks \ASMAA\ shows a fair agreement with the static points, the latter
having however a
tendency to lie a bit too high. With the EH normalization condition
[$x=1$ in eq.~\NORMCP], the situation would be much
worse, since the static points would be raised by about 30\%. Of course
this pattern
has to be checked at higher values of $\beta$\ \foot{ The authors of
ref.~\ralex\ claim that the static \FB\  decreases sensibly with increasing
$\beta$.}.

\vskip 1cm
\noindent{\bf Acknowledgements}

We wish to thank Guido Martinelli, Carlotta Pittori, Chris Sachrajda, Rainer
Sommer and Massimo Testa for many useful discussions.
\vfill

\appendix{A}{ }
This appendix is devoted to a sketch of the steps needed
to go from eq.~\IUNETDEUX\ to eq.~\RESULTILAT. Let us study
only $\CI_1^{latt}$, the analysis for $\CI_2^{latt}$ being similar.
After the integrations over $p_0$
and $k_0$, we get:

\def\Z{Z{\scriptstyle{( k)}}} \def\ZT{Z{\scriptstyle{( k+{ p\over \tau} )}}}
\def\A{A{\scriptstyle{( {p\over \tau})}}}
 \def\ZP{Z{\scriptstyle{({ p\over \tau})}}}
\def\AT{A{\scriptstyle {( k+{ p\over \tau} )}}}
\def\THETA{\Theta (1-R)}
\def\ZU{Z_1{\scriptstyle{( k)}}}
\eqn\INTEGPK{\eqalign{
\CI_1&^{latt}=           \cr
&\int\limits_{-\pi\tau}^{+\pi\tau} {d^3{\bf p}\over (2\pi)^3}
{\A -\ZP \over \A \big( {1\over \ZP}-\ZP\big)}\left(\ZP\right)^\tau
\int\limits_{-\pi}^{+\pi} {d^3{\bf k}\over (2\pi)^3} {\ZU \over \AT}
{4 \over {1\over \ZT} -\ZP\ZU}\times               \cr
&\qquad\bigg\{
{1\over {1 \over \ZU} -\ZU}\  {1-\left(\ZU\right)^\tau \over 1- \ZU} \cr
&\qquad -\qquad{\AT-\ZT \over {1\over \ZT}-\ZT} \ {1\over {\ZP\over \ZT} -\ZU}
\THETA {1-(R)^\tau \over 1-R}\cr
&\qquad -\qquad{\AT-\ZT \over {1\over \ZU}-\ZU} \ {1\over {\ZP\over \ZT} -\ZU}
{1\over \ZU\ZT}
     \THETA {1-(R)^\tau \over 1-R}\cr
&\qquad +\qquad{\AT-\ZT \over {1\over \ZU}-\ZU} \ {1\over
\ZT-\ZP\ZU}\times\cr &{\qquad \qquad \qquad \qquad \qquad \qquad \left(
 {1-\left(\ZU\right)^\tau \over 1- \ZU}- \THETA {1-(R)^\tau
\over 1-R}\right)\bigg\}\ . } \cr
}}
where
\eqn\DEF{\eqalign{
A{\scriptstyle{(k)}}\ &=\ 1\ +\ \sum\nolimits_j\left(1-\cos(k_j)\right)\cr
\ZU \ &=\ A{\scriptstyle{(k)}}\ -\ \sqrt { A{\scriptstyle{(k)}}^2-1}\cr
\Z \ &=\ 1+ B{\scriptstyle{(k)}}\ -\ \sqrt {(1+ B{\scriptstyle{(k)}})^2-1}\cr
B{\scriptstyle{(k)}}\ &=\ {\sum_j\sin^2(k_j) +(1-A{\scriptstyle{(k)}})^2
\over 2 A{\scriptstyle{(k)}}}\cr
R\ &=\ {\ZT \over \ZP}\cr }}
and $\Theta $ is the usual Heaviside distribution.

We are interested in the
behaviour of $\CI_1^{latt}$ when $\tau \to \infty$. It is
possible to show that the integration over $\bf p$ gives no problem,
essentially
because it is protected by the exponentially damping factor $\ZP^\tau$.
To illustrate the method we have used, we shall not consider the whole
expression \INTEGPK, which is rather long, but a
simpler quantity; this offers however, a degree of mathematical
complexity that is similar to that of the integrals we encounter
in the actual computation, namely:

\eqn\J{ \CJ\ =\  \int\limits_{-\pi}^{+\pi} {d^3{\bf k}\over (2\pi)^3} {1 \over
{1\over \ZT} -\ZP\ZU}   {1\over {1\over \ZU} -\ZU}\  {1-\big(\ZU\big)^\tau
\over
 1- \ZU}\ \ .}
The problem about eq.~\J\ is that, since the integral
diverges when $\tau \to \infty$, we cannot exchange the limit and
the integration. To study the behaviour of $\CJ$ as  $\tau \to \infty$,
it is convenient to first add and subtract from the integrand its value at
$p=0$; $\CJ$ is then decomposed into a sum $\CJ_1\ +\ \CJ_2$ with:
\def\ZX{Z{\scriptstyle{({k\over \tau})}}} \def\ZTX{Z{\scriptstyle{({k\over
\tau}+{ p\over \tau} )}}}
\def\ZUX{Z_1{\scriptstyle{({ k\over \tau})}}}

\eqn\CJS{\eqalign{
&\CJ_1 =  \int\limits_{-\pi}^{+\pi}
{d^3{\bf k}\over (2\pi)^3} \ {1 \over {1\over \Z} -\ZU}\
{1\over {1\over \ZU} -\ZU}\  {1-\big(\ZU\big)^\tau \over 1- \ZU}\cr
&\eqalign{{\cal J}_2 =  \int\limits_{-\pi\tau}^{+\pi\tau}
{d^3{\bf k}\over (2\pi)^3} &{1\over \tau^3}{{1\over \ZX}-{1\over \ZTX}-\ZUX
(1-\ZP)
 \over\big({1\over \ZTX} -\ZP\ZUX\big)
  \big({1\over \ZX} -\ZUX\big)}\times\cr
&{1\over {1\over \ZUX} -\ZUX}  {1-\big(\ZUX\big)^\tau \over
1- \ZUX}\ \ .\cr}\cr}}
 Let us call $J$ the integrand in $\CJ_2$. From \DEF\ we see that the
quantities $Z$ and $Z_1$ are
smaller than 1; thus we have the bound:
\eqn\MAJ{
\vert J \vert \ \le \ {1\over \tau^3} {\vert{1\over \ZX}-{1\over \ZTX}\vert +
(1-\ZP)
 \over\bigg({1\over \ZTX} -\ZUX\bigg)
  \bigg({1\over \ZX} -\ZUX\bigg)}\
{1\over {1\over \ZUX} -\ZUX}  {1-\big(\ZUX\big)^\tau \over
1- \ZUX}\ \ .}
 From \DEF, after some work, it is possible to show that there exists a
set of
 constants $\alpha_j $
such that: \eqn\BOUNDS{
\eqalign{\vert {1\over  Z(q)}-{1\over Z(q')} \vert \ &\le \ \alpha_1
\vert \mid q \mid -\mid q' \mid \vert \cr
{1 \over {1 \over Z(q)}-Z_1(q')} \ &\le \ {\alpha_2 \over \mid q \mid +
\mid q' \mid }\cr
{1-\big(Z_1(q)\big)^\tau \over 1-Z_1(q)} \ &\le
 \ \alpha_3 {1-e^{-\alpha_4 \mid q\mid}
\over \mid q \mid}\cr
(1-Z(q))\ &\le \ \alpha_5 \mid q\mid\ \ . \cr
}}
It then follows  that the absolute value of $J$ is bounded by an integrable
function of $k$ and, by Lebesgue's theorem, that we
can exchange the
limit and the integration to get:
 \def\K{\mid k \mid}
\eqn\LIMJ{\lim_{\tau \to \infty} \CJ_2\ =\
\int\limits_{-\infty}^{+\infty} {d^3{\bf k} \over (2\pi)^3}
{1-e^{-\K} \over 4 \K^3}{\K -\mid k+p \mid - \mid p \mid \over \K +
 \mid k+p \mid + \mid p \mid }\ \ .}
When the actual integral \INTEGPK\ is analysed in this way, it turns out
that the terms that are analogous
to $\CJ_2$ are responsible for the appearance of the constant $C_1$ in
eq.~\RESULTILAT. The
subtracted terms (analog to $\CJ_1$) give:
\def\AK{A{\scriptstyle {( k )}}}
\eqn\SUB{
\eqalign{\int\limits_{-\pi}^{+\pi} {d^3{\bf k} \over (2\pi)^3}
{\ZU\over \AK }&{2 \over {1\over \Z}- \ZU} {1\over \ZU -\Z}\cr
&\left( {\AK -\Z \over {1\over \Z}-\Z}\ {1-\big(\Z\big)^\tau \over 1-\Z}-
{\AK -\ZU \over {1\over \ZU}-\ZU}\ {1-\big(\ZU\big)^\tau \over
1-\ZU}\right)\ \ ,\cr}} which can be rewritten in a simpler form if we
introduce
an integration over $k_0$:
\eqn\FIN{\intpipi {e^{ik_0}-1-\sum_j\left(1-\cos(k_j)\right)\over
2 \Delta_1 \Delta_2}\ {1- e^{ik_0\tau} \over 1-e^{ik_0}}} leading to
the first term in the r.h.s. of \RESULTILAT.

\appendix{B}{ }
In this appendix, we shall derive formula \ZCHMOINSZCL. The zeroth-order graph
\new\  (a) is equal to $3/\pi^2 t^{-3}$,  and together with
eqs.~\RESULTILAT\ and \RESULTIH, we find from our definitions
\eqn\UNO{\eqalign{
e^{M_ct}& {{P^{tree}(t)\ +\ P_{stat}^c(t)}\over {P^{tree}_L(t)\ +\
P_{L}^c(t)}} - 1\cr &\eqalign{\quad{\textstyle{ \
\sim \ \atop t
\rightarrow \infty }}\ &{4\over 3}g^2\Biggl\{
\intpipi {e^{ik_0}-1-\sum\limits_j (1-\cos(k_j))
  \over 2 \Delta_1(k) \Delta_2(k_0,{\bf k})}\
{1-e^{ik_0\tau} \over 1-e^{ik_0}}\cr
& -\mu^\epsilon \int\limits_{-\infty}^{+\infty}
{d^{4-\epsilon}{\bf k} \over (2\pi)^{4-\epsilon}}
{e^{ik_0t} -1 \over{ { 1\over  2}} (k^2)^2}\ \Biggr\}\ \ .}\cr}}
If we use the identity:
\eqn\IDENTITY{
{1\over \Delta_2}\ =\ {1\over 4\Delta_1}\ +\ {\sum\limits_\lambda
\sin^4({k_\lambda \over 2}) \over \Delta_1 \Delta_2}\ -\ {\Delta_1 \over
\Delta_2}\ \ ,}
the lattice integral in eq.~\UNO\ can be rewritten as:
\eqn\DEUXIO{\eqalign{ I_L&=
\intpipi
{e^{ik_0}-1-\sum_j(1-\cos(k_j)) \over 2 \Delta_1(k) \Delta_2(k_0,{\bf k})}
\ {1-e^{ik_0\tau} \over1-e^{ik_0}}\cr &=
\intpipi {1-e^{ik_0\tau} \over 8\Delta_1^2}+{1 \over 2}\left(
{\sum\limits_\lambda \sin^4({k_\lambda \over 2}) \over \Delta_1^2
\Delta_2}-{1\over \Delta_2}\right) \left( 1- e^{ik_0\tau}\right)\cr
+&\qquad {1 \over 2}\left( {\sum\limits_j (1-\cos(k_j))
 \over \Delta_1 \Delta_2}-
{1\over
\Delta_2(0,{\bf k})}\right){1-e^{ik_0\tau}\over 1-e^{ik_0}}+
\int\limits_{-\pi}^{+\pi}{d^3{\bf k}\over (2\pi)^3} {1\over 2 \Delta_2(0,{\bf
k})}\ \ .\cr}}
 When $\tau$ is large, the factor $e^{ik_0\tau}$ gives no contribution
in the second and the third terms in the second equality of eq.
\DEUXIO. In the first term, the denominator can be exponentiated and
one arrives at some combination of Bessel functions.
This method was discussed in appendix C of ref. \LIN\ and leads to the formula:
\eqn\IUN{
\intpipi {1-e^{ik_0\tau}\over 8 \Delta_1^2}\ {\textstyle{ \
\sim \ \atop \tau
\rightarrow \infty }}\
 {1\over 8\pi^2}\left(
2\ln\left({\tau \over 2}\right) + F_{0000}+\gamma_E \right)\ \ .}
The computation of the continuum integral in eq.~\UNO\ is
straightforward and gives:
\eqn\IC{\eqalign{
I_C\ &=\ \mu^\epsilon \int\limits_{-\infty}^{+\infty}
{d^{4-\epsilon}{\bf k} \over (2\pi)^{4-\epsilon}} {e^{ik_0t} -1 \over{
{ 1\over  2}} (k^2)^2}\cr
&=\ {1\over 8 \pi^2}\left({2\over \epsilon }+
\ln(4\pi) +2\ln\left( {\mu t \over
2}\right) +\gamma_E\right)\ \ .\cr}}
 From eqs.~\DEUXIO, \IUN\ and \IC, one finally obtains eq.~\ZCHMOINSZCL.
\listrefs
\listfigs
\bye